\documentclass[12pt]{article}

\usepackage[margin=1in]{geometry} 
\usepackage{setspace}
\usepackage{lipsum}               
\usepackage[authoryear]{natbib}
\usepackage{titlesec}

\titleformat{\subsection}[hang]{\bfseries\itshape\large}{\thesubsection}{1em}{}

\titleformat{\subsubsection}[hang]{\itshape\normalsize}{\thesubsubsection}{1em}{}

\usepackage{longtable}
\usepackage[acronym]{glossaries}
\newacronym{msr}{MSR}{Mars Sample Return}
\newacronym{nasa}{NASA}{National Aeronautics and Space Administration}
\newacronym{esa}{ESA}{European Space Agency}
\newacronym{mav}{MAV}{Mars Ascent Vehicle}
\newacronym{ero}{ERO}{Earth Return Orbiter}
\newacronym{ccrs}{CCRS}{Capture, Containment, and Return System}
\newacronym{ees}{EES}{Earth Entry System}
\newacronym{bpp}{BPP}{Backward Planetary Protection}
\newacronym{os}{OS}{Orbiting Sample}
\newacronym{jpl}{JPL}{Jet Propulsion Laboratory}
\newacronym{uq}{UQ}{Uncertainty Quantification}
\newacronym{lhs}{LHS}{Latin Hypercube Sampling}
\newacronym{qoi}{QoI}{Quantity of Interest}
\newacronym{mle}{MLE}{Maximum Likelihood Estimation}
\newacronym{ig}{IG}{Inverse Gamma}
\newacronym{nllh}{NLLH}{Negative Log Likelihood}
\newacronym{nig}{NIG}{Normal Inverse Gamma}
\newacronym{am}{AM}{Adaptive Metropolis}
\newacronym{ap}{AP}{Adaptive Proposal}
\newacronym{dram}{DRAM}{Delayed Rejection Adaptive Metropolis}
\newacronym{g}{G}{gravitational force}
\newacronym{mala}{MALA}{Metropolis-adjusted Langevin algorithm}
\newacronym{gcd}{GCD}{Geweke Convergence Diagnostic}
\newacronym{gp}{GP}{Gaussian Process}
\newacronym{reml}{REML}{\textit{Restricted Maximum Likelihood}}
\newacronym{cv}{CV}{Cross Validation}
\newacronym{mspe}{MSPE}{mean squared prediction error}
\newacronym{rmse}{RMSE}{root mean squared error}
\newacronym{ci}{CIs}{credible intervals}
\newacronym{evt}{EVT}{Extreme Value Theory}
\newacronym{doe}{DoE}{Design of Experiments}
\newacronym{alr}{ALR}{Active Learning Reliability}
\newacronym{pdf}{pdf}{probability density function}
\newacronym{map}{MAP}{maximum a posteriori}
\newacronym{iid}{iid}{independently and identically distributed}
\newacronym{cdf}{cdf}{cumulative distribution function}
\newacronym{gpd}{GPD}{Generalized Pareto Distribution}
\newacronym{pc}{PC-Kriging}{Polynomial-Chaos Kriging}
\usepackage{amsmath}
\usepackage{amsthm}
\usepackage{amsfonts}
\usepackage{bbm}
\usepackage{amssymb}
\usepackage{geometry}
\usepackage{color}
\usepackage{float}
\usepackage[ruled, vlined]{algorithm2e}
\usepackage{graphicx}
\usepackage{subcaption}
\usepackage{booktabs}
\usepackage{bm}
\usepackage{dsfont}
\usepackage{mathtools}
\usepackage{lscape}
\usepackage{array}
\newcommand{\blind}{0}

\begin{document}



\def\spacingset#1{\renewcommand{\baselinestretch}%
{#1}\small\normalsize} \spacingset{1}


\if0\blind
{
\title{A Case Study on Quantifying Reliability under Extreme Risk Constraints in Space Missions}
\author{Dawn L. Sanderson\thanks{Corresponding Author: Department of Mathematics and Statistics, Air Force Institute of Technology, dawn.sanderson@au.af.edu, USA}  \and Amy Braverman\thanks{Jet Propulsion Laboratory, California Institute of Technology, amy.braverman@jpl.nasa.gov, USA} \and Giuseppe Cataldo\thanks{National Aeronautics and Space Administration, giuseppe.cataldo@nasa.gov, USA} \and Ralph C. Smith\thanks{Department of Mathematics, North Carolina State University, rsmith@ncsu.edu, USA}
\and Richard L. Smith\thanks{Department of Statistics and Operations Research, University of North Carolina at Chapel Hill, rls@email.unc.edu, USA}}
\date{}
  \maketitle
} \fi

\if1\blind
{
    \title{A Case Study on Quantifying Reliability under Extreme Risk Constraints in Space Missions}
\date{}
\maketitle
} \fi

\bigskip
\begin{abstract}
In this paper, we employ a Bayesian approach to uncertainty quantification of computer simulations used to assess the probability of rare events. As a case study, we assess the reliability of an Earth reentry capsule for sample return missions that must be able to withstand the reentry loads in order to land intact.
Our study uses Gaussian Process modeling under a Bayesian regime to analyze the reentry vehicle's resilience against operational stress. This Bayesian framework allows for a detailed probabilistic evaluation of the system's reliability, indicating our ability to verify stringent safety goals of rare events with a $0.999999$ of probability of success. The findings underscore the effectiveness of Bayesian methods for complex uncertainty quantification analyses of computer simulations, providing valuable insights for computational reliability analysis in a risk-averse setting.
\end{abstract}

\noindent%
{\it Keywords:}  rare event, computer experiments, kriging
\vfill

\newpage
\spacingset{2} 

\section{Problem description}
\label{sec:intro}
In reliability analysis, quantifying the probability of a rare event, such as a system failure occurring with probability less than one in a million, poses a significant challenge. When the system is evaluated via an expensive computer simulation, and when only limited data are available to characterize input uncertainty, traditional methods of uncertainty quantification (UQ) may not be adequate. This issue is further complicated in settings where failure is defined in terms of exceeding an extreme threshold, meaning that both interpolation and extrapolation become critical.

In such cases, surrogate modeling is typically used to emulate the computer code and provide predictions at untested input values. \gls{gp} modeling has become a standard approach in this context, offering both flexibility and interpretability \citep{sacks1989,stein1999,rasmussen2006}. These surrogate models are often paired with simulation-based techniques to estimate quantities of interest, such as failure probabilities, under a probabilistic framework. However, when input data are sparse, the uncertainty in the input distributions themselves must be incorporated into the analysis. Failing to do so can produce misleading estimates, particularly when the rare event lies in the tail of the response distribution.

Bayesian methods provide a natural framework for incorporating both parameter and model uncertainty in this setting. Prior work has established Bayesian approaches to reliability analysis \citep{sank_maha2011,sank_maha2015} and Bayesian calibration of computer models \citep{kennedy2001,Ohagan2006,Bayarri2007}. These techniques allow for formal updating of uncertain model parameters and for coherent propagation of uncertainty through the simulation process. More recent work has also addressed practical challenges in surrogate modeling for complex systems, including computational efficiency, heteroskedasticity, and high-dimensional input spaces \citep{Gramacy2020,Binois2018}.

Despite these developments, relatively few studies have addressed the combination of three factors critical in extreme reliability assessment: (1) sparse or limited empirical data on input distributions; (2) surrogate modeling of simulator outputs via GPs, where the spatial correlation structure itself is uncertain; and (3) estimation of very low tail probabilities via posterior simulation. In particular, while GP modeling is well established, the spatial range parameters are often fixed via maximum likelihood, rather than treated probabilistically and integrated into the overall uncertainty. The impact of this modeling choice can be especially pronounced when predicting values in the tails, where no simulator runs exist and where extrapolation is necessary.

This paper presents a fully Bayesian approach for estimating rare event probabilities in simulation-based reliability analysis. The input variables are assumed to follow known distributional forms, with parameters estimated from small-sample data using Bayesian methods. The output of a deterministic computer simulation is modeled using a Gaussian process (GP), with prior distributions placed on the spatial range parameters to account for correlation uncertainty in the surrogate. Cross-validation is used to tune hyperparameters and regularization penalties. Posterior distributions for all model components are then propagated through to the output prediction, and the final quantity of interest, the probability that the simulator output exceeds a critical threshold (which we call probability of failure $(\text{P}_{f})$), is estimated via posterior simulation. A safety-critical system with one-in-a-million reliability requirements is used to illustrate the method.

In the following section, an overview of the data utilized is provided. In Section \ref{sec:methods}, we outline the methodology applied during our analysis, as well as the results of the analysis in this case study. Section \ref{sec:conclusion} provides a recommendations for employing the results of our work in practice and suggestions for future work based on our findings.

\section{Data collection and preparation} 
\label{sec:data}
The data used in this study originate from a set of structural simulations intended to evaluate the preliminary design response of the Earth reentry system of a space sample return mission under high-impact loading~\citep{Cataldo_2025a}. Engineers used a finite element model to simulate the capsule's Earth reentry. The simulations focused on how variation in material properties, specifically strength and stiffness parameters, would affect the peak acceleration of the sample container experienced under nominal conditions. Empirical measurements of these properties were obtained through laboratory testing of two composite materials (in varying directions), and were then used to define the uncertain input variables for the model.

More specifically, engineers identified 15 input variables comprising Young's modulus, shear modulus, and compressive and tensile strength in varying directions for two different materials, IM7 (a carbon fiber) and Kevlar \citep{Carpenter2021}. Experimental tests executed by the engineers provided initial observations for each of these 15 input variables. The number of observations for each variable ranges from 3 to 12. Previous \gls{uq} and reliability analysis work \citep{Naresh2018, Fitt2019} as well as typical engineering practice suggest fitting two-parameter Weibull distributions to the strength variables and Normal distributions to the Modulus variables. Because of this precedence, as well as our small sample sizes and the need for extrapolation, we assume Weibull and Normal distributions, respectively. We use the indicators $X0001 - X0015$ to identify the input variables moving forward as can be seen in Table \ref{tab:variables}.

\begin{table}[H]
\centering
\footnotesize
\setlength\tabcolsep{4pt} 
\renewcommand{\arraystretch}{0.9} 
\begin{tabular}{ccccc}
\toprule
\textbf{Indicator} & \textbf{Parameter} & \textbf{Description} & \textbf{Num of Obs} & \textbf{Distributional Assumption} \\
\midrule
X0001 & $IM7-E_a,E_b,E_c$ & Young's Modulus & 8 & Normal \\
X0002 & $IM7-G_{ab},G_{bc},G_{ca}$ & Shear Modulus & 4 & Normal \\
X0003 & $IM7-X_c$ & Compressive Strength (0 Deg) & 3 & Weibull \\
X0004 & $IM7-X_t$ & Tensile Strength (0 Deg) & 4 & Weibull \\
X0005 & $IM7-Y_c$ & Compressive Strength (90 Deg) & 3 & Weibull \\
X0006 & $IM7-Y_t$ & Tensile Strength (90 Deg) & 4 & Weibull \\
X0007 & $IM7-S_c$ & Shear Strength & 4 & Weibull \\
X0008 & $Kv-E_a,E_c$ & Young's Modulus (0 Deg) & 4 & Normal \\
X0009 & $Kv-E_b$ & Young's Modulus (90 Deg) & 4 & Normal \\
X0010 & $Kv-G_{ab},G_{bc},G_{ca}$ & Shear Modulus & 4 & Normal \\
X0011 & $Kv-X_c$ & Compressive Strength (0 Deg) & 12 & Weibull \\
X0012 & $Kv-X_t$ & Tensile Strength (0 Deg) & 4 & Weibull \\
X0013 & $Kv-Y_c$ & Compressive Strength (90 Deg) & 11 & Weibull \\
X0014 & $Kv-Y_t$ & Tensile Strength (90 Deg) & 4 & Weibull \\
X0015 & $Kv-S_c$ & Shear Strength & 4 & Weibull \\
\bottomrule
\end{tabular}
\caption{Description of input variables}
\label{tab:variables}
\end{table}

After determining the parameters of the respective Normal and Weibull distributions for the 15 input variables using \gls{mle}, the engineers next generated 25 samples of the input variables using \gls{lhs}. Using these 25 samples, the engineers then ran 25 simulations of an LS-DYNA Earth entry vehicle's system-level impact model and extracted the peak acceleration as the quantity of interest \citep{LS-DYNA,SLIM_2022,SLIM_2023}. In the context of our analysis, we consider the 25 realization of the peak acceleration as our output variable of interest. With the particulars of the data in mind, we next review the methods applied throughout our analysis. 

\section{Analysis and interpretation}
\label{sec:methods}
\subsection{Methods}
 In the following sections, we outline the various methodologies employed in our analysis. A direct \gls{mle} approach is not appropriate because of the high dimension of the variable space and the low sample size; though we compare a functional regularized  \gls{reml} approach to our preferred Bayesian method, which we favor due to its effective ability to incorporate uncertainty in the parameters. We begin with a consideration of the priors applied within the Bayesian analysis of our input variables.

\subsubsection{Prior considerations}\label{subsec:weibull_prior}

We have 15 input variables, each represented by either a Weibull or a Normal distribution; our goal with the Bayesian approach is to generate posterior distributions of the input variables' parameters. For all the input variables, regardless of distributional assumptions, we compare the frequentist estimates provided by the \gls{reml} values and the mean of the posterior distributions generated by several priors within a Bayesian framework. That is, we assume a prior density on the unknown parameters $\pi_0(\psi)$ (here $\psi=(\alpha,\beta)$ for Weibull variables or $\psi=(\mu,\sigma^2)$ for Normal variables). Then, by applying Bayes' Rule, we have $\pi(\psi \mid X_{1},\hdots,X_{n}) = \frac{\pi_0(\psi)\prod_{i=1}^n{f(X_i\mid \psi)}}{\int \pi_0(\psi^{\prime})\prod_{i=1}^n{f(X_i\mid \psi^{\prime})}d\psi^{\prime}}$, where the distribution $\pi(\psi \mid X_{1},\hdots,X_{n})$ is known as the posterior distribution and reflects the updated knowledge of the parameters conditional on the data \citep{Gelman2013}. 

The prior distribution is meant to reflect the prior knowledge one has in regard to the parameters. The degree of certainty surrounding these parameters can be controlled by the type of prior used in the analysis. The three priors we consider are a flat prior, which is non-informative, assigning equal probability to all parameter values; Jeffreys' prior, which is also non-informative but scale-invariant and is defined by the square root of the determinant of the Fisher information matrix \citep{Jeffreys1939}; and the conjugate prior, in which the prior and posterior distributions are part of the same probability family \citep{Gelman2013} For the Weibull distribution we have an Inverse Gamma conjugate prior and for the Normal distribution, the Normal Inverse Gamma conjugate prior. While using certain priors, such as the conjugate prior, allows us to know the form of the posterior distribution, it is not always a straightforward task to generate a posterior distribution. In the next section, we discuss the use of an \gls{am} algorithm \citep{Haario2001} that we implemented throughout our analysis in order to generate our posterior distributions.

\subsubsection{Adaptive Metropolis (AM) algorithm}\label{subsec:AM_alg}
Once a prior distribution has been chosen, we look to generate values from the posterior distribution, $\pi(\boldsymbol{\psi}|X_{1} \hdots X_{n})$. One method of accomplishing this when the above distribution is intractable is by use of a Metropolis algorithm \citep{Metropolis1953}. This is an algorithm that can be used to obtain random samples from a probability distribution using a general \textit{symmetric} proposal distribution where there is an associated accept/reject rate for the proposal distribution. 


While this is a well-known, often-used algorithm, one issue typically cited is the difficulty in choosing a proposal distribution. The choice of proposal distribution greatly affects the speed of the algorithm and the acceptance probability, which is typically desired to be between $20-50\%$ \citep{Gelman1996}. A feasible solution exists in the application of adaptive algorithms ``which use the history of the process in order to `tune' the proposal distribution suitably" \citep{Haario2001}. \citet{Haario2001} developed an \gls{am} algorithm that adapts continuously to the target distribution and is based on the original Metropolis algorithm \citep{Metropolis1953} and its modifications as well as the \gls{ap} algorithm given in \citet{Haario1999}. 

The \gls{ap} algorithm uses a Gaussian proposal distribution centered on the current state with the covariance calculated from a fixed finite number of previous states \citep{Haario1999}. The change to the \gls{am} algorithm is that the covariance of the proposal distribution is calculated using all the previous states \citep{Haario2001}. For further details on the specifics of the \gls{am} Algorithm and its implementation, we refer the reader to \citet{Haario2001}.

Different variations of the Metropolis and Metropolis Hastings algorithms \citep{hastings1970} were considered, such as the \gls{dram} \citep{haario2006}, the Hamiltonian Monte Carlo \citep{neal2011}, or the \gls{mala} \citep{roberts2002}; however, for the balance in speed of computation and accuracy of algorithm results as well as the ease of implementation, we chose the \gls{am} algorithm. Another consideration in regards to the algorithm and its output was the number of iterations to apply and the amount of burn-in, if any, to remove. To make this determination, we relied on trace plots, running average plots, and the \gls{gcd} \citep{geweke1992}. Now that we have discussed the \gls{am} algorithm, we turn our attention to the subjects of Gaussian processes and spatial statistics in relation to our modeling approach.


\subsubsection{Gaussian processes and spatial statistics}\label{subsec:GP_spatialstats}

First we consider the basic approach to modeling complex processes using computer simulations. Given a selection of input variables $\mathbf{X}=(X_1,...,X_k)$, an output variable, $Z=h(\mathbf{X})$, is produced by the computer code. In many instances, the reason for generating such output variables is to then use the results for making further predictions. \citet{sacks1989} discusses the use of stochastic processes in modeling the response and making predictions while \citet{Gramacy2020} expand upon this by using Bayesian methods in the calibration of computer models to improve the prediction process. 

For our study, we model the output, $Z$, using a \gls{gp}, a widely-used tool for dealing with spatially structured data \citep{rasmussen2006}. The \gls{gp} provides a flexible, nonparametric approach, which is particularly apt at capturing spatial correlations \citep{cressie1993}. We define the \gls{gp} model as $Z \sim N_n(X\beta,\Sigma)$, where Z represents the vector of simulation outputs, $X\beta$ denotes the mean response as a function of parameters $\beta$ ($X$ represents the design matrix of the \gls{gp}, not the set of input variables previously defined as $\mathbf{X}$), and $\Sigma$ is the covariance matrix capturing the dependence structure among the data.

In \gls{gp}s, the covariance between any two observations depends on the input locations corresponding to these observations. That is, we let $\Sigma=\alpha V(\mathbf{\theta})$, where $\alpha>0$ is a scale parameter and $V$ is a function of the spatial range parameters $\mathbf{\theta}$ and the euclidean distances between observations.
Considered a popular choice for such applications, we use the squared exponential function for $V$ \citep{rasmussen2006,Hadji2019}. This form allows for an anisotropic covariance structure by using different spatial range parameters $\theta_k$, which scale the distances along each input dimension $k$, thus capturing varying influences of the inputs. We define $V$ by its individual entries:
\begin{equation}\label{eq:vform}
     v_{ij}=\exp{\bigg(-\sum_{k=1}^K \frac{(x_{ik}-x_{jk})^2}{\exp(\theta_k)^2}\bigg)}
\end{equation}
where $x_{ik}$ and $x_{jk}$ are the $k$-th input variables of observations $i$ and $j$. 

In order to estimate our spatial range parameters, we will again apply a Bayesian framework; however, we begin with outlining the MLE approach as presented by Stein (1999). Given our definition of $Z$ as above, we have:
\begin{equation}\label{Zpdf}
   f(Z|X\beta,\Sigma) = (2\pi)^{-n/2}|\Sigma|^{-1/2}\exp{\bigg(-\frac{1}{2}[Z-X\beta]^T\Sigma^{-1}[Z-X\beta] \bigg)} 
\end{equation}
This produces the negative log likelihood:
$$l(\beta,\alpha,\theta)=\frac{n}{2} \log{(2\pi)} +\frac{n}{2}\log{\alpha} +\frac{1}{2}\log{|V(\theta)|}+\frac{1}{2\alpha}[Z-X\beta]^TV(\theta)^{-1}[Z-X\beta] $$
Using the least squares estimator of $\beta$ based on the covariance matrix $V(\theta)$:\\ 
\begin{equation}\label{bhat}
    \hat{\beta}(\theta)=(X^TV(\theta)^{-1}X)^{-1}X^TV(\theta)^{-1}Z
\end{equation}
and defining:\\
$H=V(\theta)^{-1}-V(\theta)^{-1}X(X^TV(\theta)^{-1}X)^{-1}X^TV(\theta)^{-1}$ and $G^2(\theta)=Z^THZ$, we then have:
$$l(\hat{\beta}(\theta),\alpha,\theta)=\frac{n}{2}\log{(2\pi)}+\frac{n}{2}\log{\alpha} +\frac{1}{2}\log{|V(\theta)|}+\frac{1}{2\alpha}G^2(\theta)$$
Finally, we can minimize this analytically with respect to $\alpha$, resulting in: $\hat{\alpha}(\theta) = \frac{G^2(\theta)}{n}$, giving us:
$$l^*(\theta)=l(\hat{\beta}(\theta),\hat{\alpha}(\theta),\theta)=\frac{n}{2}\log{(2\pi)} +\frac{n}{2}\log{\frac{G^2(\theta)}{n}}+ \frac{1}{2}\log{|V(\theta)|}+\frac{n}{2}$$
The above quantity is often called the \textit{profile negative log likelihood}.

Another option in terms of likelihoods that arises in the realm of spatial statistics is the \textit{restricted maximum likelihood} or REML approach, which was first introduced by Patterson and Thompson (1971). The idea behind the REML method is to separate the two part estimation problem, that of estimating the linear model and the estimation of the covariance structure. This is done by considering the likelihood function of the contrasts; that is, let $W=A^TZ$ be a vector of $n-q$ linearly independent contrasts, then $W \sim N(0,A^T\Sigma A)$. This gives the pdf:
$$f(W \mid A,\Sigma)= (2\pi)^{-(n-q)/2}|A^T \Sigma A|^{-1/2}\exp{\left(-\frac{1}{2}W^T|A^t\Sigma A|^{-1}W \right)}$$
with the negative log likelihood:
$$l_w(\alpha,\theta)=\big(\frac{n-q}{2}\big)\log{(2\pi \alpha)} +\frac{1}{2}\log{|A^T V(\theta) A|}+\frac{1}{2}W^T|A^T V(\theta)A|^{-1}W$$
Following the calculations of Patterson and Thompson (1971) and suggestions made by Harville (1977), the above likelihood can be simplified to:
$$l_w(\alpha,\theta) =\big(\frac{n-q}{2}\big)\log{(2\pi \alpha)} -\frac{1}{2}\log{X^TX}+\frac{1}{2}\log{|X^TV(\theta)^{-1}X|}+\frac{1}{2}\log{|V(\theta)|}+\frac{1}{2\alpha}G^2(\theta) $$
Finally, minimizing with respect to $\alpha$, results in $\tilde{\alpha}=\frac{G^2(\theta)}{n-q}$, giving us:
\begin{align}
\begin{split}
     l_w^*(\theta)= \frac{n-q}{2}\log{2\pi}+\frac{n-q}{2}\log{\frac{G^2(\theta)}{n-q}}&-\frac{1}{2}\log{|X^TX|}\\
   &+\frac{1}{2}\log{|X^TV(\theta)^{-1}X|}+\frac{1}{2}\log{|V(\theta)|}+\frac{n-q}{2}
\end{split}
\label{REML}
\end{align}
Therefore, $l_w^*(\theta)$ is what we will refer to as the \textit{REML negative log likelihood}.

Before moving on to the Bayesian approach, we include a brief mention of a regularized REML method. With some models, there might exist stabilization issues with the estimates; or, estimation problems can arise when the number of parameters exceeds the number of data points. The regularized REML approach works to prevent these issues by including a penalty term to the likelihood function. We will apply a Ridge-type regularization where our \textit{regularized REML negative log-likelihood} takes the form: 
\begin{equation}
    l_R(\theta)=l_w^*(\theta)+\lambda \sum_{k=1}^K(\theta_k-\bar{\theta})^2
    \label{eq:regREML}
\end{equation}
Here, the penalty term is $\lambda$ and the L2 penalty function tends to shrink the spatial range parameters towards their mean value.

Finally, we turn to the Bayesian approach where a prior, $\frac{\pi(\theta)}{\alpha}$, is placed on the spatial range parameters. Using our defined pdf from Equation \ref{Zpdf} and defining $\hat{\beta}$ as in \ref{bhat} we have the following posterior distribution:
\begin{align*}
    \pi(\beta,\alpha,\theta \mid Z) \propto \frac{\pi(\theta)}{\alpha}\alpha^{-n/2}|V(\theta)|^{-1/2} &\cdot \exp{\bigg[-\frac{G^2(\theta)}{2\alpha}\bigg]}\\
    &\cdot \exp{\bigg[-\frac{1}{2\alpha}(\beta-\hat{\beta})^TX^TV(\theta)^{-1}X(\beta-\hat{\beta})\bigg]}
\end{align*}
Integrating out with respect to $\beta$ and $\alpha$ we have:
$$\pi(\theta \mid Z) \propto \pi(\theta)|V(\theta)|^{-1/2}G^2(\theta)^\frac{q-n}{2}|X^TV(\theta)^{-1}X|^{-1/2}$$
Which gives us the \textit{Bayesian negative log likelihood} of:
\begin{equation}
    l_B(\theta)=-\log{\pi(\theta)}+\frac{1}{2}\log{|V(\theta)|}+\frac{n-q}{2}\log{G^2(\theta)}+\frac{1}{2}\log{|X^TV(\theta)^{-1}X|}
    \label{eq:bayes_like}
\end{equation}

While we presented the derivation of several versions of applicable negative log-likelihood equations, we note here that we utilize the regularized REML as the frequentist version in comparison to the Bayesian approach. In the next section, we discuss the use of these varying methods, to include the Bayesian method with the prior $\theta_{1:K} \sim N(\tau,\nu^2)$. We use Leave One Out \gls{cv} twice, first to make an appropriate choice for the penalty term in the regularized \gls{reml} case and again when choosing the hyperparameters $(\tau, \nu^2)$ in the Bayesian case.

Once we have generated the posterior distributions for $\theta_{1:K}$, we can then predict new values from the computer simulations using a kriging approach. Kriging, a geostatistical interpolation technique, provides optimal, unbiased predictions by utilizing the spatial correlation structure inferred by the Gaussian process model \citep{cressie1993,stein1999}. We apply ordinary kriging to the problem, aiming to predict a new value $z_0$ with mean $x_0^T \beta$, variance $\sigma_0^2$, and covariance $\text{Cov}(z_0,\mathbf{Z})=\phi$. Following the Lagrange Multiplier approach with Lagrange multiplier $\gamma$, we consider predictors of the form $\hat{z}_0=\gamma^T Z$ subject to the constraint $\gamma^T X = x_0^T$. Thus, we end up minimizing the \gls{mspe}: $E[(z_0-\hat{z_0})^2]=\sigma_0^2-2\gamma^T \phi +\gamma^T \Sigma \gamma$ subject to the above constraint. This gives us the predictor and \gls{mspe}:
\begin{equation}\label{eq:pred}
    \hat{z}_0=(x_0-X^T\Sigma \phi)\hat{\beta}+\phi^T \Sigma^{-1}\mathbf{Z}
\end{equation}
\begin{equation}\label{eq:mspe}
     E[(z_0-\hat{z_0})^2] = \sigma_0^2 \phi^T\Sigma^{-1}\phi +(x_0-X^T\Sigma^{-1}\phi)^T(X^T \Sigma^{-1}X)^{-1}(x_0-X^T \Sigma^{-1}\phi)
\end{equation}

After we have generated the posterior distributions for the parameters of the Normal and Weibull variables as well as for the spatial range parameters, we are able to predict the peak acceleration of the sample container. A distinct advantage of the Bayesian approach to generate the spatial range parameters, as opposed to estimating the spatial process using \gls{reml} values for $\theta_{1:K}$, is that we are accounting for the error in the estimation of the spatial range parameters. This provides a more robust quantification of the underlying uncertainty within the model. With these elements in place, we can then simulate the posterior distribution of the probability of exceeding 3000 Gs with the given information.

\subsubsection{Cross validation in a Bayesian setting}\label{subsec:CV}

We mentioned in Section \ref{subsec:GP_spatialstats} that we would be using \gls{cv} in two instances. In the first scenario, when choosing the penalty term for the regularized \gls{reml} case, our \gls{cv} follows the methods outlined by \citet{hastie2009}. For both uses of \gls{cv}, we employ a squared error loss function. The algorithm used for the first CV is outlined in Algorithm \ref{alg:CV1}. The input values for the algorithm are the input variables, $\mathbf{X}_{n\times K}$, the output variable, $\mathbf{Z}_{n \times 1}$, and the selected potential values for $\lambda$, $\mathbf{\Lambda}_{Q \times 1}$. The output of the algorithm is the vector of \gls{cv} scores for each value of $\lambda$ used, $\mathbf{L}_{Q \times 1}$.

With the second use of \gls{cv}, when choosing the hyperparameters in the Bayesian case, we follow the same methods but must account for the fact that we are dealing with posterior distributions of the $\theta_k$ values. In order to accommodate this addition, we adjust Algorithm \ref{alg:CV1} by applying the kriging step to every realization from the posteriors of $\theta_{1:K}$ (after removing burn-in) then calculate the squared error loss for each realization. Finally, we use the mean of the squared error loss across all of these realizations as the \gls{cv} score for each potential set of hyperparameters. This update can be seen in Algorithm \ref{alg:CV2}. We have the same input and output values as in Algorithm \ref{alg:CV1} except we use $\mathbf{P}_{Q \times 2}$ to represent the potential hyperparameter values instead of having a vector for the choice of $\lambda$. We also identify $T$ as the number of \gls{am} algorithm outputs and $b$ as the number of burn-in iterations removed.

Now that we have discussed all the elements that are used for generating the various posterior distributions, we move to the final step which is the simulation of the $\text{P}_{f}$.

\singlespacing
\begin{algorithm}[H]
\footnotesize
\DontPrintSemicolon 
\KwIn{$\mathbf{X}_{n\times K}$, $\mathbf{Z}_{n \times 1}$, $\mathbf{\Lambda}_{Q \times 1}$, $n$, $Q$}
\KwOut{The set of CV values $\mathbf{L}_{Q \times 1}$ where $L_q=\sum_{i=1}^n(z_{0i}-\hat{z}_{0i}^{(q)})^2$}
\BlankLine
\For{$q \in 1,2,\hdots,Q$}{
  Set value of $\lambda$ from $\mathbf{\Lambda}$ \;
  \For{$i \in 1,2,\hdots,n$}{
  Omit entry $i$ from $\mathbf{X}$ and $\mathbf{Z}$, set $n_i=n-1$ and $x_0$ as omitted vector from $\mathbf{X}$.\;
  Run optimization routine using Equation \ref{eq:regREML} to generate estimates for $\theta_{1:15}$.\;
  Perform kriging steps to generate $\hat{z}_{0i}^{(q)}$. \;
  } 
  Calculate $L_q = \sum_{i=1}^n(z_{0i}-\hat{z}_{0i}^{(q)})^2$
} 
\Return{Result: $\mathbf{L}_{Q \times 1}$}\;
\caption{Cross Validation for Penalty Term $\lambda$\label{alg:CV1}}
\end{algorithm}

\singlespacing
\begin{algorithm}[H]
\footnotesize
\DontPrintSemicolon 
\KwIn{$\mathbf{X}_{n\times K}$, $\mathbf{Z}_{n \times 1}$, $\mathbf{P}_{Q \times 2}$, $n$, $Q$, $T$, $b$}
\KwOut{The set of CV values $\mathbf{L}^*_{Q \times 1}$ where $L_q^*=\frac{1}{T-b}\sum_{j=1}^{T-b}L_{qj}$}
\BlankLine
\For{$q \in 1,2,\hdots,Q$}{
  Set value of $\tau$ and $\nu^2$ from $\mathbf{P}$ \;
  \For{$i \in 1,2,\hdots,n$}{
  Omit entry $i$ from $\mathbf{X}$ and $\mathbf{Z}$, set $n_i=n-1$ and $x_0$ as omitted vector from $\mathbf{X}$.\;
  Run \gls{am} Algorithm using Equation \ref{eq:bayes_like} to generate posteriors: $\Theta_{15 \times T}$.\;
  remove burn-in, $b$, from posterior results. \;
  \For{$j \in 1,2,\hdots,T-b$}{
    Perform kriging steps to generate $\hat{z}_{0i_j}^{(q)}$.\;
  } 
  } 
  \For{$j \in 1,2,\hdots,T-b$}{
     Calculate $L_{qj} = \sum_{i=1}^n(z_{0i}-\hat{z}_{0i_j}^{(q)})^2$
  }
 Calculate $L^*_q=\frac{1}{T-b}\sum_{j=1}^{T-b}L_{qj}$
} 
\Return{Result: $\mathbf{L}^*_{Q \times 1}$}\;
\caption{Cross Validation for Hyperparameters $\tau$ and $\nu^2$\label{alg:CV2}}
\end{algorithm}
\doublespacing

\subsubsection{Simulation of probability of failure $(\text{P}_{f})$}\label{subsec:sim_failure}
Our end goal is to calculate the probability that the peak acceleration exceeds 3000 Gs, we will call this $z_{crit}$. In order to do this, we run a simulation based on Algorithm \ref{alg:sim}, fixing large $N$ and $M$. For each iteration of the simulation, we sample from the posterior distributions of parameters for our Weibull and Normal variables respectively in order to then generate a new set of input values $\mathbf{s}_0$. Additionally, we sample from the posterior distributions of the spatial range parameters, $\theta_{1:K}$ as input for the \gls{uq} model. Finally, we calculate the exceedance probability based on our previous GP assumption that $Z \sim N_n(X\beta,\Sigma)$. What results is a sample of probabilities that can be treated as the posterior distribution of the probability of exceeding 3000 Gs.

\singlespacing
\begin{algorithm}[H]
\footnotesize
\DontPrintSemicolon 
\KwIn{Values: $N$, $M$, and $K$. Posterior distributions: $\pi(\alpha,\beta)$, $\pi(\mu,\sigma^2)$, and $\pi(\theta_k)$ for $k \in 1,\hdots,K$.}
\KwOut{The set of values $\hat{p}_{crit}^{(i)}$, $i=1,\hdots,N$, which is treated as a sample from the posterior distribution of $p_{crit}$.}
\BlankLine
\For{$i \in 1,2,\hdots,N$}{
  \For{$k \in 1,2,\hdots,K$}{
  Select values $\alpha_k,\beta_k,\mu_k,\sigma_k$ from respective posterior distributions of the k$^{th}$ input variable.\;
  Select values $\theta_k$ from the respective posterior distributions of the k$^{th}$ spatial range parameter.\;
  \textit{These selections represent one sample from the posterior distribution.}\;
  } 
    \For{$j \in 1,2,\hdots,M$}{
        \For{$k \in 1,2,\hdots,K$}{
      Simulate one value of the Normal or Weibull distribution corresponding to the k$^{th}$ input variable.\textit{This is our trial set of inputs $\mathbf{s}_0$}.\;
      }
      Use the \gls{uq} model to obtain the mean predictor $\hat{z}_0$ and its R\gls{mspe}, $S_0$, at predictor $\mathbf{s}_0$.\;
      Calculate Exceedance probability: $1-\Phi(\frac{z_{crit}-\hat{z}_0}{S_0})$
    } 
  Average over $j=1,\hdots,M$ to obtain one estimate $\hat{p}_{crit}^{(i)}$ for the probability of exceeding $z_{crit}$.\;
} 
\Return{Result: $\hat{p}_{crit}^{(i)}, i=1,\hdots,N$}\;
\caption{$\text{P}_{f}$ Simulation\label{alg:sim}}
\end{algorithm}
\doublespacing

Having outlined the methodological framework and the statistical procedures used in this study, we now turn to the application of these methods. This includes applying our Gaussian process model and the associated kriging methodology to the data generated from the computer simulation. By employing the aforementioned techniques, we aim to illuminate the predictive capabilities of our model, demonstrating its practical relevance and validity. The following section thus presents the findings from this application, detailing the results and discussing their implications for our research question.

\subsection{Application and Results}\label{sec:app&res}
The overall analysis was completed in several steps, the first of which was to estimate the parameters for the Weibull and Normal distributions of the respective input variables. Next we estimated the spatial range parameters for the spatial model as well as evaluated the performance of our \gls{uq} model. Finally, we used Algorithm \ref{alg:sim} to compute the distribution of the $\text{P}_{f}$. All analysis was completed using R Software \citep{r2022}.

\subsubsection{Input variable analysis}\label{subsec:input_analysis}
Given the distributional assumptions on our 15 variables, we find the \gls{reml} values for $\mu$, $\sigma$ and $\alpha$, $\beta$, respectively; additionally, we apply the \gls{am} algorithm using all three of the aforementioned prior distributions (dependent on the distributional assumption). The \gls{reml} values give us an initial estimate on the parameters of the distributions. Allowing for different priors indicates how influential the choice of prior is on the posterior distribution. Additionally, when comparing the posterior means to the \gls{reml} values we can check the robustness of the results.

The following description of the settings used for the \gls{am} algorithm follow the notation provide by \citet{Haario2001}. For every application of the \gls{am} algorithm unless otherwise noted, we set the total number of simulations to: $t=100,000$; we also set the following values: $s_d=\frac{(2.4)^2}{d}$ (where $d$ is the dimension of the sample parameter space), $\epsilon=0.0001$, $t_0=10,000$, $t_1 = 10$, as well as $t_2=100$ where $t_2$ is the interval between updates of our output parameter matrix. Therefore, we end up with an output parameter matrix of dimension $\frac{t}{t_2} \times d=1,000 \times d$. The number of total simulations was chosen after some initial runs at varying values; the choice of $t=100,000$ was based on the trace plots and running average plots. The values for $t_0, t_1$, and $t_2$ were based on the speed of the algorithm as well as our confidence in the initial covariance matrix (in regards to $t_0$). After establishing the above settings of the \gls{am} algorithm, we looked at trace plots and running average plots as well as the \gls{gcd} for each application of the \gls{am} algorithm in order to assess the burn-in rate for the algorithm. Taking into consideration all of these diagnostics, we removed the first 20\% of the runs as burn-in for each use of the \gls{am} algorithm unless otherwise specified.

\begin{figure}[H]
  \centering
  \begin{subfigure}{0.4\textwidth} 
    \includegraphics[width=\textwidth]{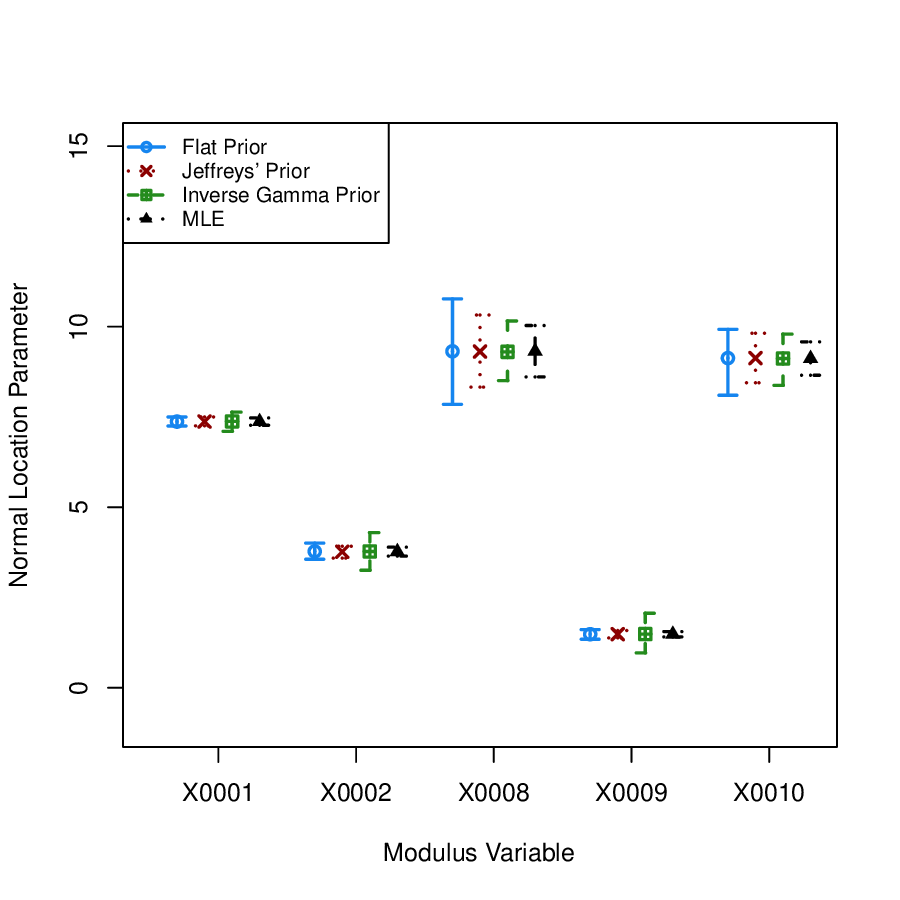}
    \caption{Plot of posterior mean with 95\% \gls{ci} for the location parameter of the Normally distributed variables.}
  \end{subfigure}%
  \hspace{0.03\textwidth} 
  \begin{subfigure}{0.4\textwidth} 
    \includegraphics[width=\textwidth]{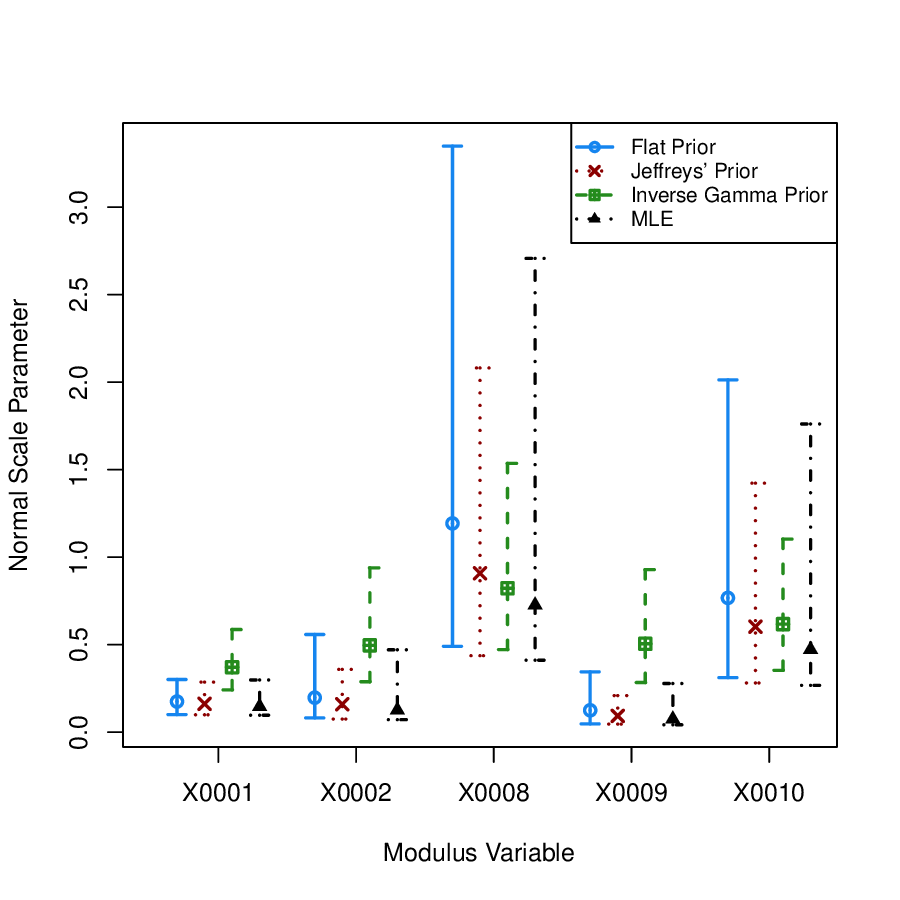}
    \caption{Plot of posterior mean with 95\% \gls{ci} for the scale parameter of the Normally distributed variables.}
  \end{subfigure}
  \caption{Posterior Distributions based on varying prior choices for the Normally distributed variables}
  \label{fig:norm_posts}
\end{figure}

The results for the normally distributed variables can be seen in Figure \ref{fig:norm_posts} where  the mean of the posterior distributions along with their 95\% \gls{ci} are plotted; additionally, the \gls{reml}s and their 95\% confidence intervals are included for comparison. As observed in Figure \ref{fig:norm_posts}(a), all the means and \gls{ci} are very similar. In Figure \ref{fig:norm_posts}(b) we see less agreement in the scale parameter values, with some variation in the mean values as well as differing lengths of \gls{ci}; however, overall, the agreement across the priors is fairly consistent with no statistically significant differences.


\begin{figure}[H]
  \centering
  \begin{subfigure}{0.4\textwidth} 
    \includegraphics[width=\textwidth]{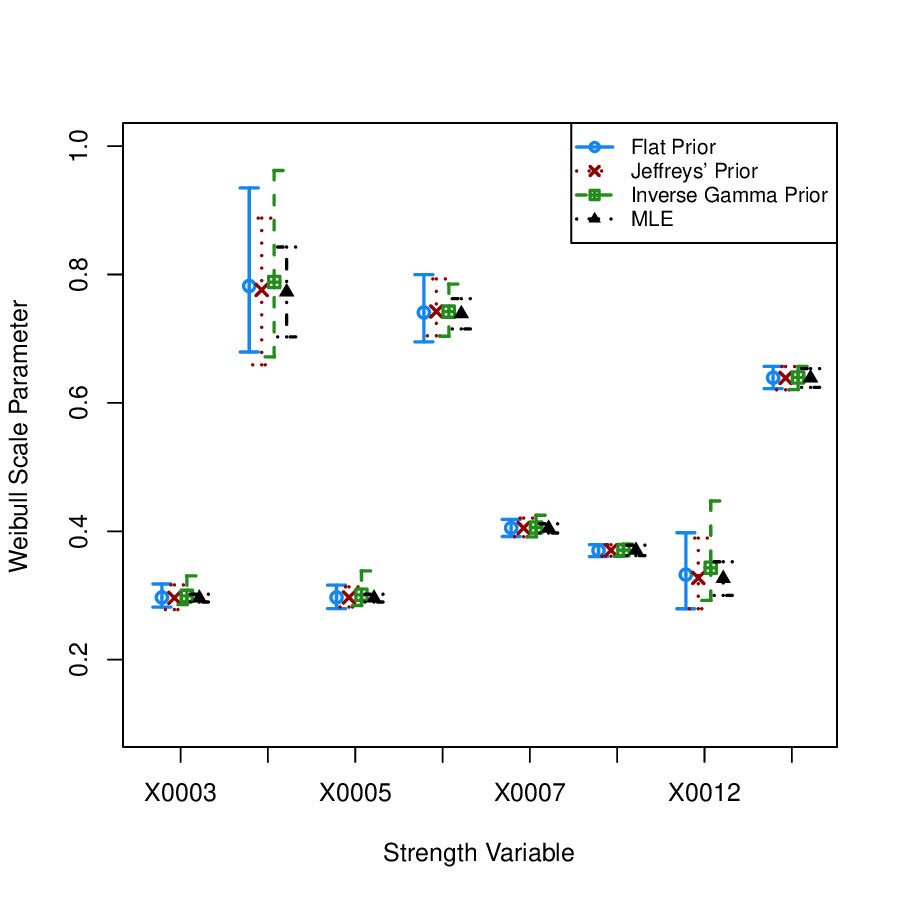}
    \caption{Plot of posterior mean with 95\% \gls{ci} for the scale parameter of the Weibull distributed variables.}
  \end{subfigure}%
  \hspace{0.03\textwidth} 
  \begin{subfigure}{0.4\textwidth} 
    \includegraphics[width=\textwidth]{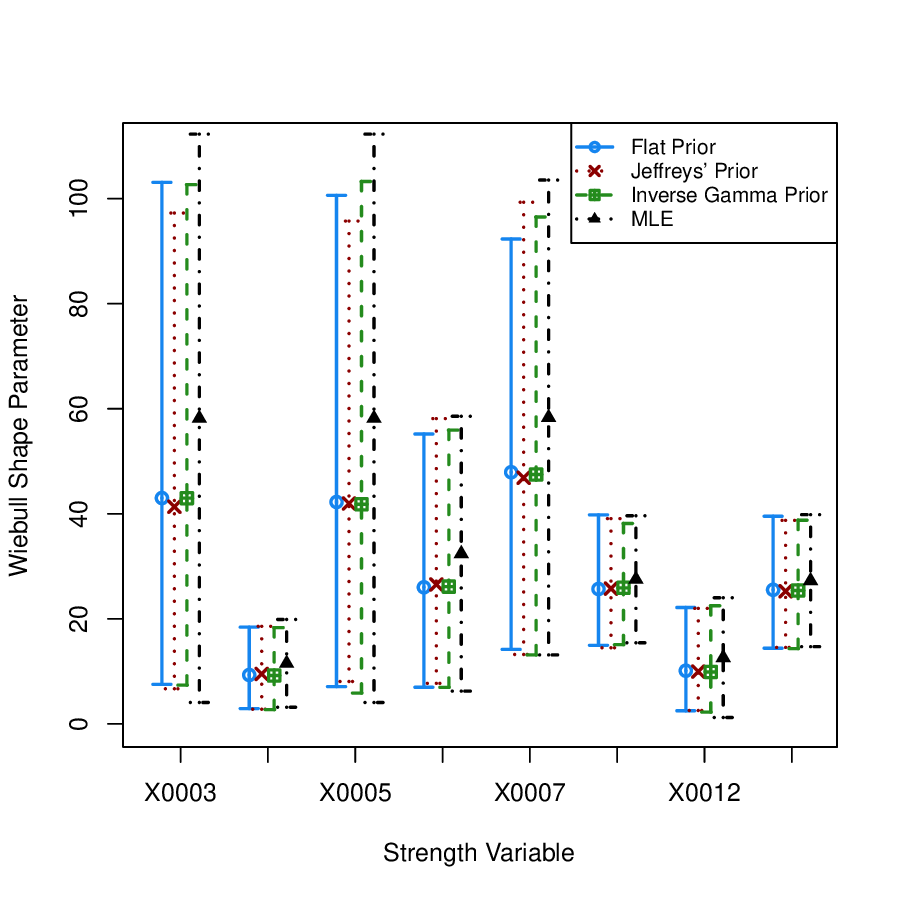}
    \caption{Plot of posterior mean with 95\% \gls{ci} for the shape parameter of the Weibull distributed variables.}
  \end{subfigure}
  \caption{Posterior Distributions based on varying prior choices for the Weibull distributed variables}
  \label{fig:weibull_posts}
\end{figure}

Next we turn to the results for the Weibull distributed variables. Again we see plots of the mean and 95\% \gls{ci} of the posterior distributions for the shape and scale parameters of the Weibull variables in Figure \ref{fig:weibull_posts}. Here we observe agreement for both the scale and shape parameters across prior choice. The consistency we see across prior choice in each instance demonstrates that the choice of prior is not overtaking the results; that is to say, the prior is not overly influential in the behavior of the posterior distributions.

Based on the results for both the Normal and Weibull variables, we use the posterior distributions that incorporated Jeffreys' prior in further analysis steps. The main reasoning behind this is that Jeffreys' prior gives more information than the flat prior (as it incorporates information about the structure of the model through the Fisher information) and does not have the extended parameter assumptions of a known shape parameter that is necessary when using a conjugate prior for the Weibull distribution.

\subsubsection{Spatial range parameters analysis}\label{subsec:srp_analysis}

Applying the \gls{gp} approach to our data, we have $\mathbf{X}=(X0001,\hdots,X0015)$ as our input variables with the previously generated \gls{lhs} as our input values for those variables. For our output variable $Z$, or the peak acceleration, we use the 25 realizations from \gls{jpl}'s work. In our analysis, for numerical stability, we re-scaled the output data by dividing all output values by a factor of 1000, making our problem determining the probability that the peak acceleration is greater than $3.0$ when the input variables are random.

\begin{figure}[H]
  \centering
  \begin{minipage}[t]{0.35\textwidth}
    \centering
    \includegraphics[width=\linewidth]{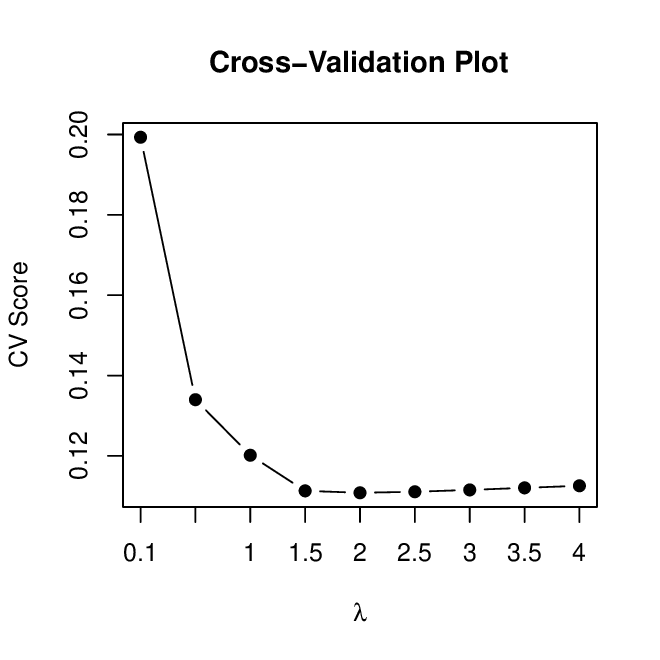}
    \subcaption{\gls{cv} for regularized REML likelihood penalty term}\label{fig:cv1}
  \end{minipage}
  \hspace{0.05\textwidth}%
  \begin{minipage}[t]{0.35\textwidth}
    \centering
    \includegraphics[width=\linewidth]{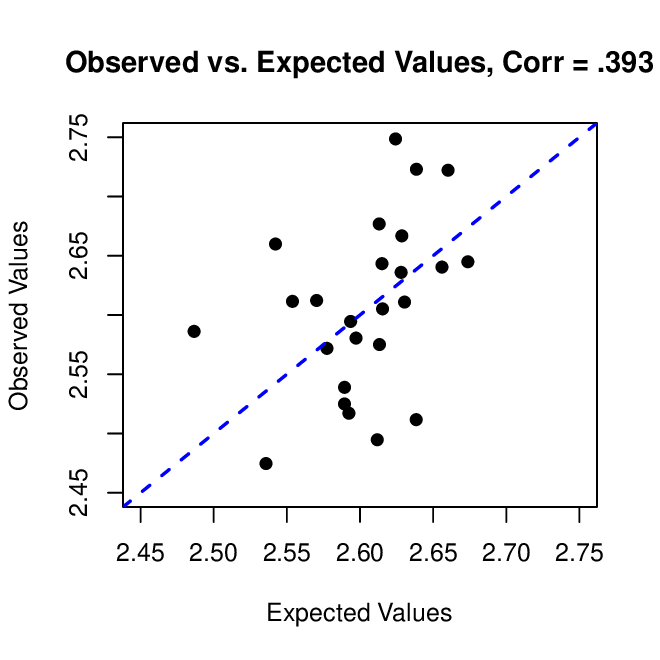}
    \subcaption{Observed vs. Expected peak acceleration where observed are calculated using kriging methods described in Section \ref{subsec:GP_spatialstats}}\label{fig:corr1}
  \end{minipage}  
  \caption{CV Results and Kriging Diagnostic Plot using Equation \ref{eq:regREML}}
  \label{fig:UQdiag1}
\end{figure}

For simplicity, we assume a constant mean response for $X\beta$ (e.g. $\mathbf{X}=\mathbf{1}_{n\times1}$). Initially, we seek to determine the \gls{reml} values for the spatial range parameters $\theta_{1:K}$ with the overarching goal of generating posterior distributions for $\theta_{1:K}$ and then drawing from those distributions during the prediction (kriging) process. Considering several values of $\lambda>0$ via \gls{cv}, we choose the value of $\lambda=2$ for our penalty term. We also completed the kriging step within the \gls{cv} analysis at this point to see how our model performs in predicting the peak acceleration. The \gls{cv} plot as well as a diagnostic plot of the model's predictive quality can be seen in Figure \ref{fig:UQdiag1}.



From Figure \ref{fig:cv1}, we see that $\lambda=2$ does provide the smallest \gls{cv} value. In Figure \ref{fig:corr1} we see that the variance of the noise is greater than that of the signal. This is supported by the weak correlation of $0.393$ and the low signal to noise ratio of $0.390$.  From this we consider that it is possible that the range of input values is too narrow; that is, we would get greater signal to noise ratio with a wider range of inputs. We also note that there are no simulated values in the critical region; that is, our $Z_{1:25}$ range from $2.474$ to $2.749$, so we have no data over $3.0$. Furthermore, kriging with a constant mean is a form of interpolation, which means it will never produce a predicted value outside the range of data and input values far from the test data set will simply be predicted back to the overall mean.

In spite of this critique, which we discuss further in the next section, we move forward with the Bayesian analysis. We use Equation \ref{eq:bayes_like} with $\theta_{1:K} \sim N(\tau,\nu^2)$ as the prior for the spatial range parameters; however, we must first determine appropriate values for $\tau$ and $\nu^2$. We start by using an Empirical Bayes approach, considering the \gls{reml} values of the Normal prior distribution for the $\theta_i$'s which are $\hat{\tau}=\bar{\theta}$ and $\hat{\nu}^2 = \frac{1}{K}\sum_{i=1}^K{(\theta_i-\hat{\tau})^2}$. Here, as we do not have the actual values of $\theta_{1:K}$, we use the values generated from Equation \ref{eq:regREML} with $\lambda=2$ to calculate $\hat{\tau}$ and $\hat{\nu}^2$. This approach suffices for the estimation of $\hat{\tau}=1.49$, as the \gls{reml} values of the $\theta_i$'s provide unbiased point estimates of their central tendency; however, the \gls{reml}-derived value of $\hat{\nu}^2$ underestimates the true variability in the $\theta_i$'s due to its reliance solely on the spread of point estimates without accounting for uncertainty in parameter estimation. This underestimation may be particularly severe here given the small sample size. To address this, we instead use the output of the Hessian matrix, $H$, from the optimization of Equation \ref{eq:regREML} and use $\hat{\nu}^2=\frac{1}{K}\sum_{i=1}^K{H_{i,i}^{-1}}$ as our estimate for $\nu^2$. This estimation accounts for the curvature of the likelihood function and better represents the uncertainty in the spatial range parameters. Using this method, we obtain $\hat{\nu}^2=0.259$, a more realistic value that incorporates the variability in both the estimated parameters and the likelihood surface itself.

\begin{figure}[H]
  \centering
  \begin{minipage}[t]{0.35\textwidth}
    \centering
    \includegraphics[width=\linewidth]{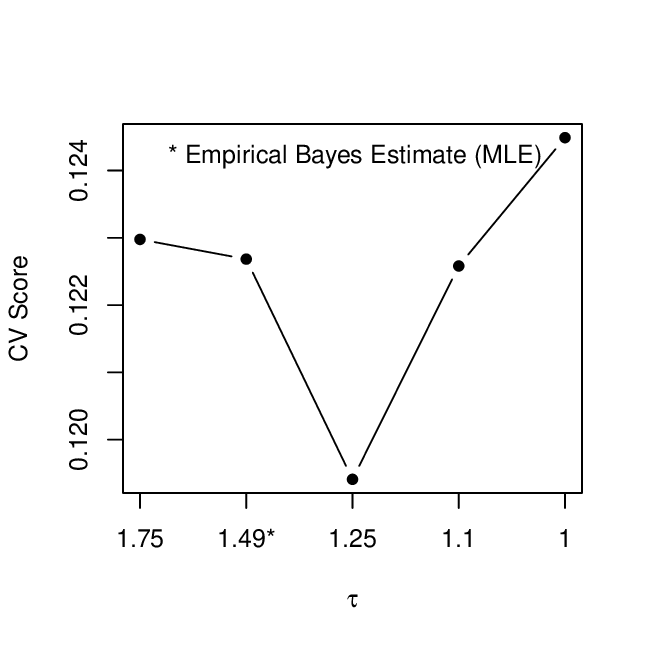}
    \subcaption{\gls{cv} for regularized REML likelihood penalty term}
  \end{minipage}
  \hspace{0.05\textwidth}%
  \begin{minipage}[t]{0.35\textwidth}
    \centering
    \includegraphics[width=\linewidth]{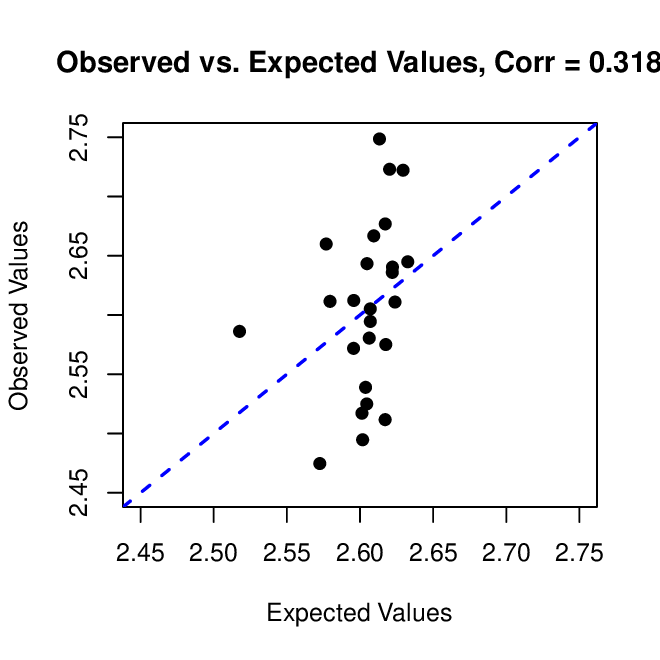}
    \subcaption{Observed vs. Expected peak acceleration where observed are calculated using kriging methods described in Section \ref{subsec:GP_spatialstats}}\label{fig:corr2}
  \end{minipage}  
  \caption{CV Results for Spatial Range Parameters and Kriging Diagnostic Plot using Equation \ref{eq:bayes_like}}
  \label{fig:theta_cv}
\end{figure}

After making the adjustment for $\hat{\nu}^2$, we then turn to \gls{cv} to determine if the MLE derived value of $\hat{\tau}=1.49$ as calculated from the output of optimizing Equation \ref{eq:regREML} is the best choice. In order to do this, for each potential value of $\tau$ under consideration, we run the adaptive algorithm $n=25$ times (as we used leave one out \gls{cv}). Because of the computing time necessary to implement leave one out \gls{cv}, we narrowed the \gls{cv} options to five potential values, $\tau =\{1.00, 1.10, 1.25, 1.49, 1.75\}$ based on some initial runs of the \gls{cv} with fewer iterations for the AM algorithm. We see the results of the more extensive (in terms of number of iteration in the \gls{am} algorithm) \gls{cv} analysis in Figure \ref{fig:theta_cv}a which shows that $\hat{\tau}=1.25$ is the best option for our estimate of the mean of our prior distribution. To check the appropriateness of the choice for $\nu^2$, we did an additional round of \gls{cv} with the same values of $\tau$ while using $\nu^2=0.5$. The \gls{cv} scores with $\nu^2=0.5$ were consistently larger than when using $\nu^2=0.259$. Because of the amount of computing time necessary to run the \gls{cv}, we did not examine different combinations of $\tau$ and $\nu^2$ simultaneously.


Once we determined the best choices for the hyperparameters of the prior distribution $(\tau = 1.25, \nu^2=0.259)$, we examine the model performance using the chosen parameters. We also compare these results to those using the regularized \gls{reml} approach with $\lambda=2$ as the penalty term. Comparing the reported correlation values in Figures \ref{fig:corr1} and \ref{fig:corr2} ($0.393$ and $0.318$, respectively), we see that the prediction using the Bayesian approach is slightly worse; however, this is to be expected given that we have allowed for more uncertainty within the choice of the spatial range parameters. 

\begin{figure}[H]
    \centering
    \includegraphics[scale=.6]{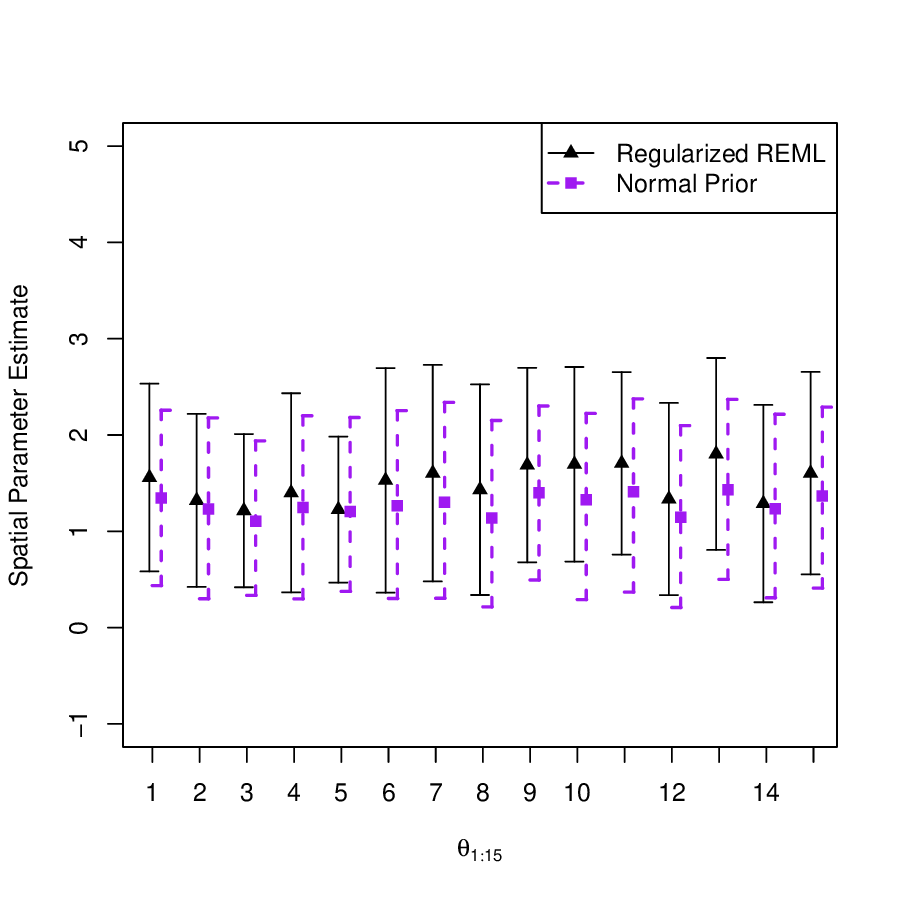}
    \caption{Comparison of \gls{reml} and Bayesian results}\label{fig:theta_comp}
\end{figure}

Finally, we examine the resulting values of the spatial range parameters from the Bayesian analysis. In Figure \ref{fig:theta_comp}, we see the \gls{reml} results from Equation \ref{eq:regREML} plotted with their 95\% confidence intervals as well as the posterior means and 95\% \gls{ci} produced by the \gls{am} algorithm. As we can see, the two methods are fairly comparable, with the posterior mean estimates from the Bayesian method pooling toward the common mean, driven by the choice of $\tau=1.25$ as the mean of the prior distribution. This regularization in the Bayesian context is to be expected given the few number of data points we have, making the influence of the prior stronger.

Now that we have posterior distributions for all of the parameters of our Normal and Weibull variables $(X0001,\hdots,X0015)$ as well as our spatial range parameters $(\theta_{1:15})$, we can move forward with our end-to-end analysis.

  
  

\subsubsection{Final simulation results}\label{subsec:fin_results}
\begin{figure}[H]
  \centering
  \begin{subfigure}{0.4\textwidth} 
    \includegraphics[width=\textwidth]{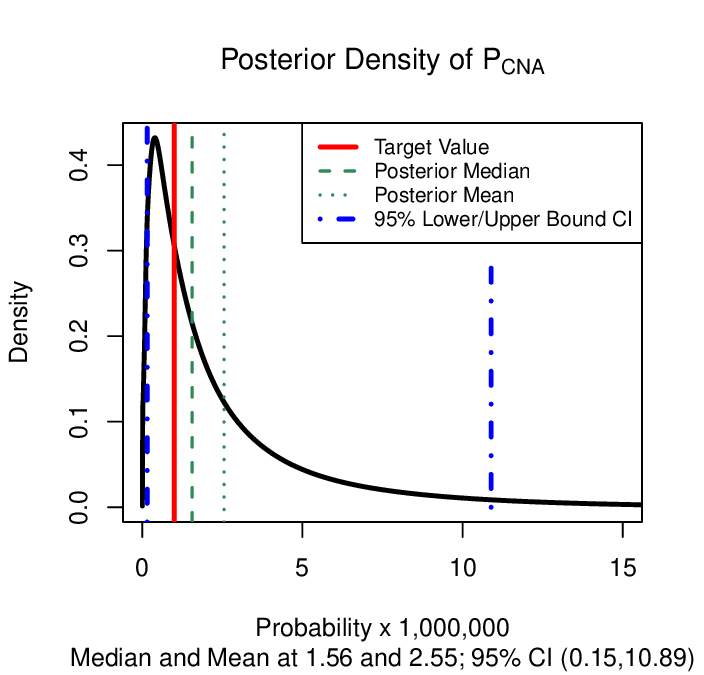}
    \caption{Plot of posterior distribution of the $\text{P}_{f}$ under Setting A}
  \end{subfigure}%
  \hspace{0.03\textwidth} 
  \begin{subfigure}{0.4\textwidth} 
    \includegraphics[width=\textwidth]{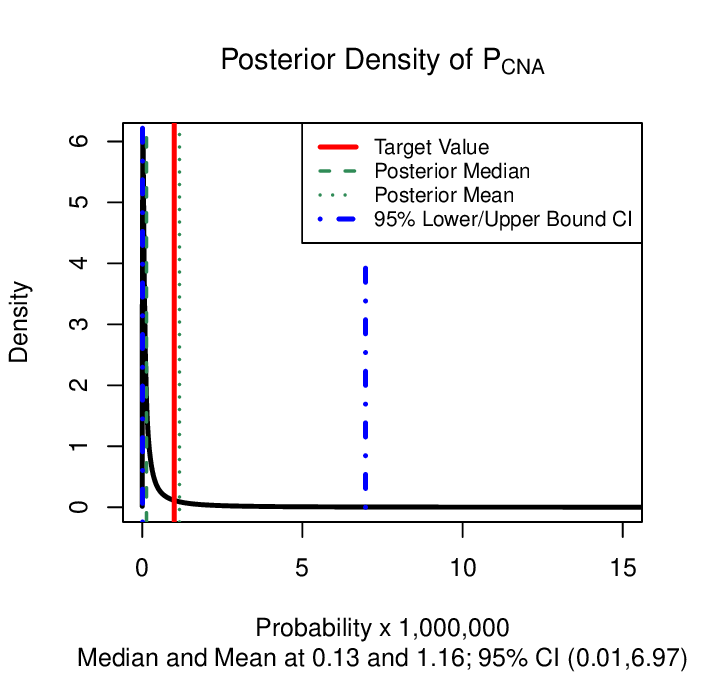}
    \caption{Plot of posterior distribution of the $\text{P}_{f}$ under Setting B}
  \end{subfigure}
  \caption{Comparison of Posterior Distributions of $\text{P}_{f}$ between settings A and B}
  \label{fig:final_graphs}
\end{figure}

We run the end-to-end simulation found in Algorithm \ref{alg:sim} using two settings. In both settings A and B, we use the posterior distributions for $\alpha,\beta,\mu,\sigma^2$ generated by applying the \gls{am} algorithm with the respective Jeffreys' priors. For the spatial range parameters $\theta_{1:15}$ in Setting A, we use the \gls{reml} results given by Equation \ref{eq:regREML} with penalty term $\lambda = 2$. And in Setting B, for the spatial range parameters $\theta_{1:15}$ we use the posteriors produced by applying the \gls{am} algorithm with Equation \ref{eq:bayes_like} and prior $\theta_k \sim N(1.25,0.259)$. In both settings, we let $N=2000$ and $M=1000$. The resulting posterior distributions of the $\text{P}_{f}$ can be seen in Figure \ref{fig:final_graphs}.

As observed in Figure \ref{fig:final_graphs}a under Setting A, both the median and mean of the posterior distribution fail to meet the target value of one, or the probability of one in a million. However, in Figure \ref{fig:final_graphs}b, the median of the posterior is well below the target value at $0.13$ and the mean only slightly exceeds the target value at $1.16$. We also see a narrower credible interval under Setting B. Thus, even though we have included a greater amount of uncertainty by varying the values of the spatial range parameters in Setting B, the $\text{P}_{f}$ appears to be reduced. A deeper dive into this result concluded that this change is due to the slight decrease in the spatial range parameter estimates and is specifically tied to the choice of prior parameters. 

While this raises some concerns, we emphasize the extensive exploration in selecting the parameter values of the prior via \gls{cv}. That is, in the context of this problem, we have incorporated empirical evidence by beginning with the \gls{reml} values for the spatial range parameters and then applying CV to choose hyperparameter values for the prior on the spatial range parameters in the Bayesian framework. Although the variability in $\theta_{1:K}$ has a strong influence on the final $\text{P}_{f}$ distribution, we have built some assurance in the results in Figure \ref{fig:final_graphs}b with the systematic approaches outlined above. The implications of these findings and potential avenues for future research are discussed in the following section.

\vspace{-.2in}
\section{Advice to practitioners}\label{sec:conclusion}

The general process outlined and applied above gives a Bayesian approach to \gls{uq} of computer simulations. Using the simulation of peak acceleration for a space sample return mission as a case study, we have demonstrated a model for providing a measurement of the $\text{P}_{f}$ for an Earth reentry capsule. By employing Bayesian methods, we have produced a probability density of the $\text{P}_{f}$ as opposed to a singular point estimate. While our method and the subsequent results have the limitations indicated below, our analysis suggests that the predominant goal of ensuring a safe landing at least equal to $99.9999\%$ may be conceivable for this particular component of a reentry capsule. Additionally, providing decision makers with \gls{ci} allows for further risk assessment and analysis in the event that the stated goal has some flexibility~\citep{CataldoInPrep}.

Several of the limitations within this analysis suggest directions for future projects. The first was mentioned in Section \ref{subsec:srp_analysis} in regard to kriging with a constant mean. While this element of the model could be modified, we believe the real issue here is the data used to generate the initial set of output values. We mentioned the engineers running the experiment used \gls{lhs} when sampling from the input variables. This is a perfectly valid method of sampling as outlined by \citet{mckay79}. However, the 25 \gls{lhs} samples fail to produce an example of the peak acceleration exceeding 3000 Gs when run through the chosen computer model. 
Although the experiment was constructed in line with well-established principles for LHS sampling, it is nevertheless problematic that we are trying to calculate an exceedance probability for 3000 Gs when no member of the simulation exceeds that level. To avoid this issue, we suggest a study on the comparison of sampling methods in respect to their impact on the $\text{P}_{f}$.

Although we believe that our method effectively exploits the data that are available, we also believe it highlights the disadvantages of working with such a limited dataset. As the initial computer model only produced 25 simulations of the peak acceleration, we were very limited in testing the robustness of our UQ model. There were multiple reasons for the number of runs conducted by \gls{jpl}, which include the cost of experiments as well as the length of time the simulation software takes. However, given the work we have done here, our counterparts at \gls{jpl} have been receptive to the following suggestion. We propose the engineers generate more simulations using their computer model, perhaps using a sampling scheme as indicated by our next planned study. This would allow us to reapply the methods used in this paper and explore further alternative techniques if we meet the same sensitivity issues in regards to the influence of the choice of spatial range parameters in the GP model.

Finally, we outlined the chain of influence the choice of the spatial range parameters has on the final probability distribution. To explore this influence further, the future study with the incorporation of additional data should consider alterations to the model. These alterations would include examining a change to the GP's covariance structure, and/or a change to the GP's mean function, as well as a larger scale change in the distributional assumptions for the output variable $Z$ (e.g., a t-distribution). Exploring these different avenues, bolstered by additional data, will allow us to hone our methodology within this particular problem set and strengthen our abilities to apply such methods to similar problems in other space missions.

In conclusion, our work shines a light on the significant potential of the Bayesian approach in \gls{uq} for computer simulations, particularly in the context of complex projects such as, for example, \gls{msr}~\citep{Cataldo_2024,Sarli24}.
The highlighted limitations and subsequent recommendations not only provide a roadmap for immediate improvements but also emphasize the interdisciplinary nature of this research, bridging the gap between statistical methodologies and engineering challenges. As we advance further into an era dominated by simulations and computational models, understanding the underpinnings of their uncertainties becomes paramount. As we move forward, augmenting our dataset and refining our methodologies will be instrumental in enhancing the reliability and precision of our computational assessments.

\newpage
{\large\bf Disclosure Statement}

The authors report there are no competing interests to declare.

\if0\blind
{
{\large\bf Acknowledgments}

The authors would like to thank their colleagues Aaron Siddens and Kevin Carpenter at the Jet Propulsion Laboratory for providing the dataset used in this manuscript.

{\large\bf Funding}

Funding provided by the National Aeronautics and Space Administration is gratefully acknowledged. Portions of the research were carried out at the Jet Propulsion Laboratory, California Institute of Technology, under a contract with the National Aeronautics and Space Administration (80NM0018D0004).

} \fi

{\large\bf Data Availability}

Due to the proprietary nature of the original engineering simulations, the dataset used in this study cannot be publicly shared. However, a representative dataset that preserves the structure and statistical properties of the original data has been created to support reproducibility. This simulated dataset, along with annotated code used for the analysis, is available at https://doi.org/10.5281/zenodo.15855903.

\newpage
\bibliography{main}


\end{document}